\documentclass{iaus}
\usepackage{graphicx}

\title[Ring galaxies and giant low surface brightness galaxies] 
{Are ring galaxies the ancestors of giant low surface brightness galaxies?}

\author[M. Mapelli et al.]   
{M. Mapelli$^1$, B. Moore$^1$, E.~Ripamonti$^2$, L.~Giordano$^1$, L.~Mayer$^{1,3}$, M.~Colpi$^4$ \and  S.~Callegari$^1$}

\affiliation{$^1$Istitute for theoretical physics, University of Zurich,\\ Winterthurerstrasse 190, CH 8057, Z\"urich, Switzerland email: {\tt mapelli@physik.unizh.ch}\\[\affilskip]
$^2$Dipartimento di Fisica e Matematica, Universit\`a dell'Insubria, \\ via Valleggio 11, I- 22100, Como, Italy\\[\affilskip]
$^3$Institute of Astronomy, ETH Z\"urich, \\ ETH Honggerberg HPF D6, CH-8093, Z\"urich, Switzerland\\[\affilskip]
$^4$Universit\`a di Milano Bicocca, Dipartimento di Fisica G. Occhialini, \\ Piazza delle Scienze 3, I-20126 Milano, Italy
}

\pubyear{2008}
\volume{xxx}  
\pagerange{xxx}
\jname{The Galaxy Disk in Cosmological Context}
\editors{J. Andersen (Chief Editor), J. Bland-Hawthorn \&{} B. Nordstr\"om, eds.}
\begin{document}

\maketitle

\begin{abstract}
Giant low surface brightness galaxies (GLSBs) have flat discs extending up to $\sim{}100$ kpc. Their formation is a puzzle for cosmological simulations in the cold dark matter scenario. We suggest that GLSBs might be the final product of the evolution of collisional ring galaxies. In fact, our simulations show that, approximately $0.5-1.5$ Gyr after the collision which led to the formation of the ring galaxy, the ring keeps expanding and fades, while the disc becomes very large ($\sim{}100$ kpc) and flat. At this stage, our simulated galaxies match many properties of GLSBs (surface brightness profile, morphology, HI spectrum and rotation curve).
\end{abstract}

\firstsection 
\section{Introduction}
The giant low surface brightness galaxies (GLSBs) are low surface brightness galaxies (LSBs) characterized by the unusually large extension of the stellar and gaseous disc (up to $\sim{}100$ kpc; \cite[Pickering et al. 1997]{Pickering+97} and references therein) and by the presence of a normal stellar bulge (\cite[Sprayberry et al. 1995]{Sprayberry+95}; \cite[Pickering et al. 1997]{Pickering+97}). Their prototype is Malin~1 (\cite[Bothun et al. 1987]{Bothun+87}). The existence of GLSBs is a puzzle for cosmology and in particular for cosmological simulations in the cold dark matter scenario. In fact, most of cosmological simulations including a baryonic component produce galactic discs which are too compact and too bulge-dominated to match the properties even of a Milky Way-like galaxy (\cite[D'Onghia et al. 2006]{donghia+06}). Thus, there is no way to form GLSBs, whose discs are flat and huge, in current cosmological simulations. Various mechanisms have been proposed for the origin of GLSBs, but none of them is able to solve completely the problem. For example, a large-scale bar can redistribute the disc matter and increase the disc scalelength (\cite[Noguchi 2001]{Noguchi01}). However, bar instabilities normally do not increase the disc scalelength by more than a factor of 2.5, which is not sufficient to produce the observed GLSBs.

In this proceeding, we show that the propagation of the ring in an old collisional ring galaxy can lead to the redistribution of  mass and angular momentum in both the stellar and gas component
out to a distance of $\sim{}100-150$ kpc from the centre of the galaxy, producing features (e.g. the surface brightness profile, the star formation, the HI emission spectra and the rotation curve) which are typical of GLSBs. 

\begin{figure}[b]
\begin{center}
 \includegraphics[width=6.4in]{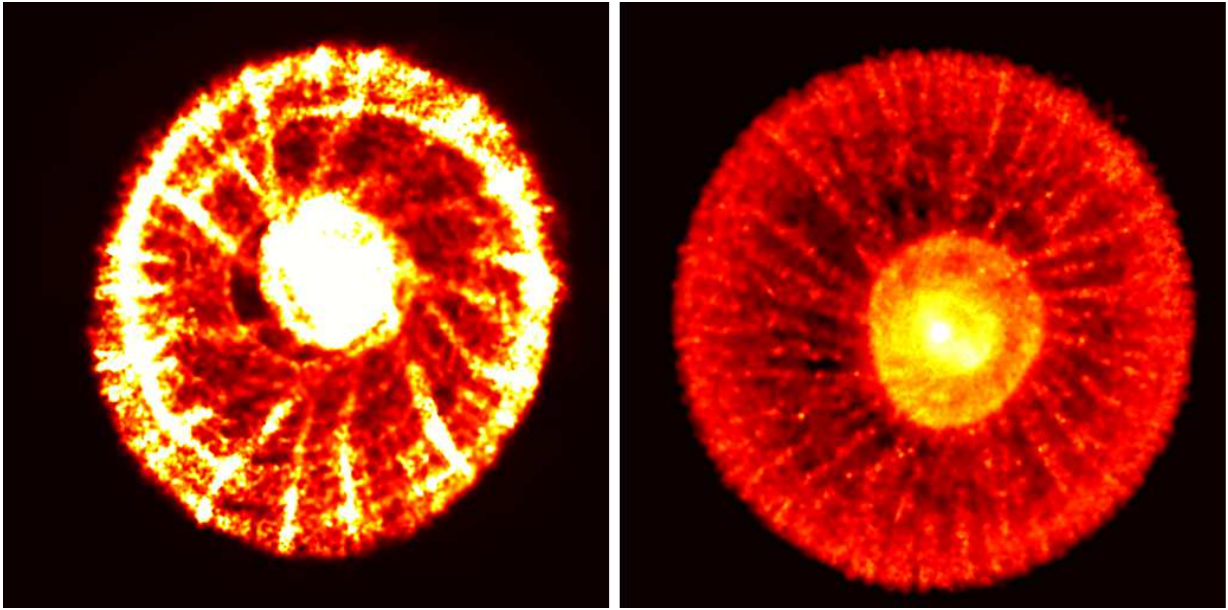} 
 \caption{Stellar density map of a simulated ring galaxy. The density is projected along the $z-$ axis. Left-hand panel: run A3 of \cite[Mapelli et al. 2008a]{Mapelli+08a}, $\sim{}100$ Myr after the galaxy interaction. The frame measures 90 kpc per edge. The density scales from 0 to 27 $M_\odot{}$ pc$^{-2}$ in linear scale. 
Right-hand panel: run C of \cite[Mapelli et al. 2008b]{Mapelli+08b}, $\sim{}1$ Gyr after the galaxy interaction. The frame measures 260 kpc. The colour coding indicates the density, projected along the $z$-axis, in logarithmic scale (from 2 to 70 $M_\odot{}$ pc$^{-2}$).  
.}
   \label{fig1}
\end{center}
\end{figure}


\begin{figure}[b]
\begin{center}
 \includegraphics[width=6.4in]{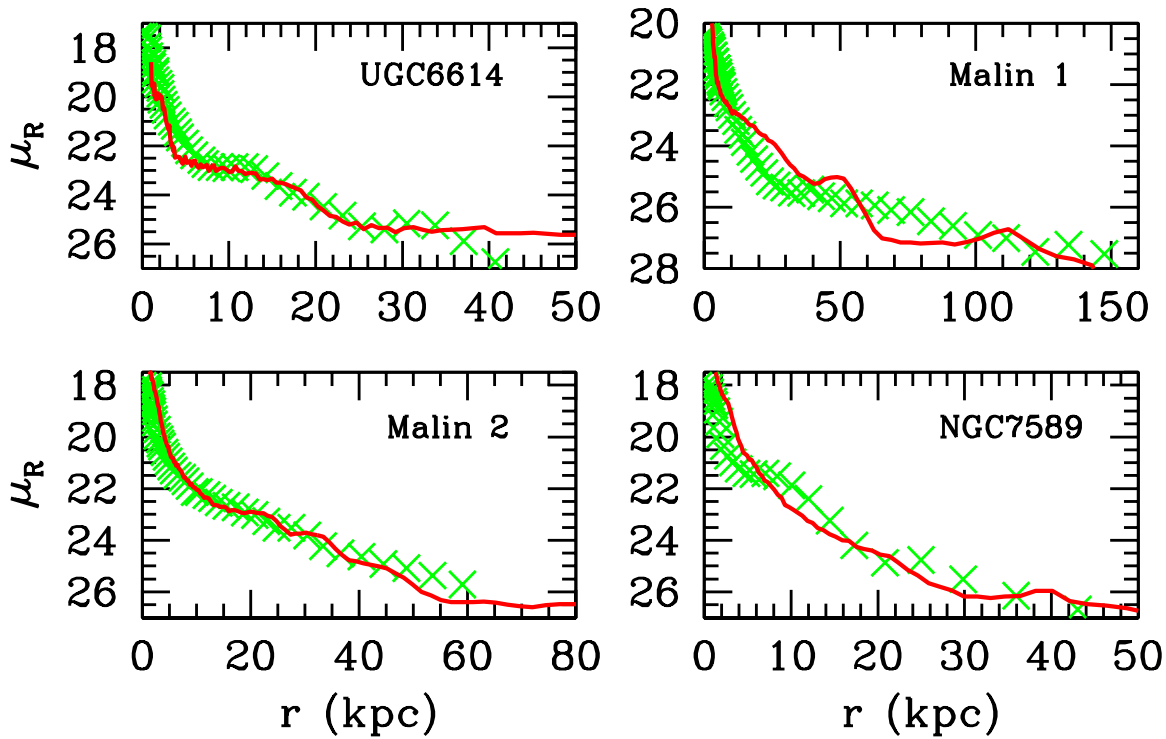} 
 \caption{$R-$band stellar surface brightness profile of the GLSB sample (in units of magnitude per arcsec$^2$). Green crosses: data points from \cite[Pickering et al. (1997)]{pickering+97} for UGC6614, Malin 2 and NGC7589, and from \cite[Moore \& Parker (2006)]{moore06} for Malin 1. The $1\,{}\sigma{}$ errors are of the same order of magnitude as the points. Red solid line: stellar surface brightness profile in $R$ magnitude derived from the simulations. From top to bottom and left to right: UGC6614 data and run C of \cite[Mapelli et al. (2008b)]{Mapelli+08b} (at time $t=0.5$ Gyr), Malin 1 data and run C of \cite[Mapelli et al. (2008b)]{Mapelli+08b} ($t=1.4$ Gyr), Malin 2 data and run B of \cite[Mapelli et al. (2008b)]{Mapelli+08b} ($t=1.0$ Gyr), NGC7589 and run A of \cite[Mapelli et al. (2008b)]{Mapelli+08b} ($t=0.5$ Gyr).}
   \label{fig2}
\end{center}
\end{figure}

\begin{figure}[b]
\begin{center}
 \includegraphics[width=6.4in]{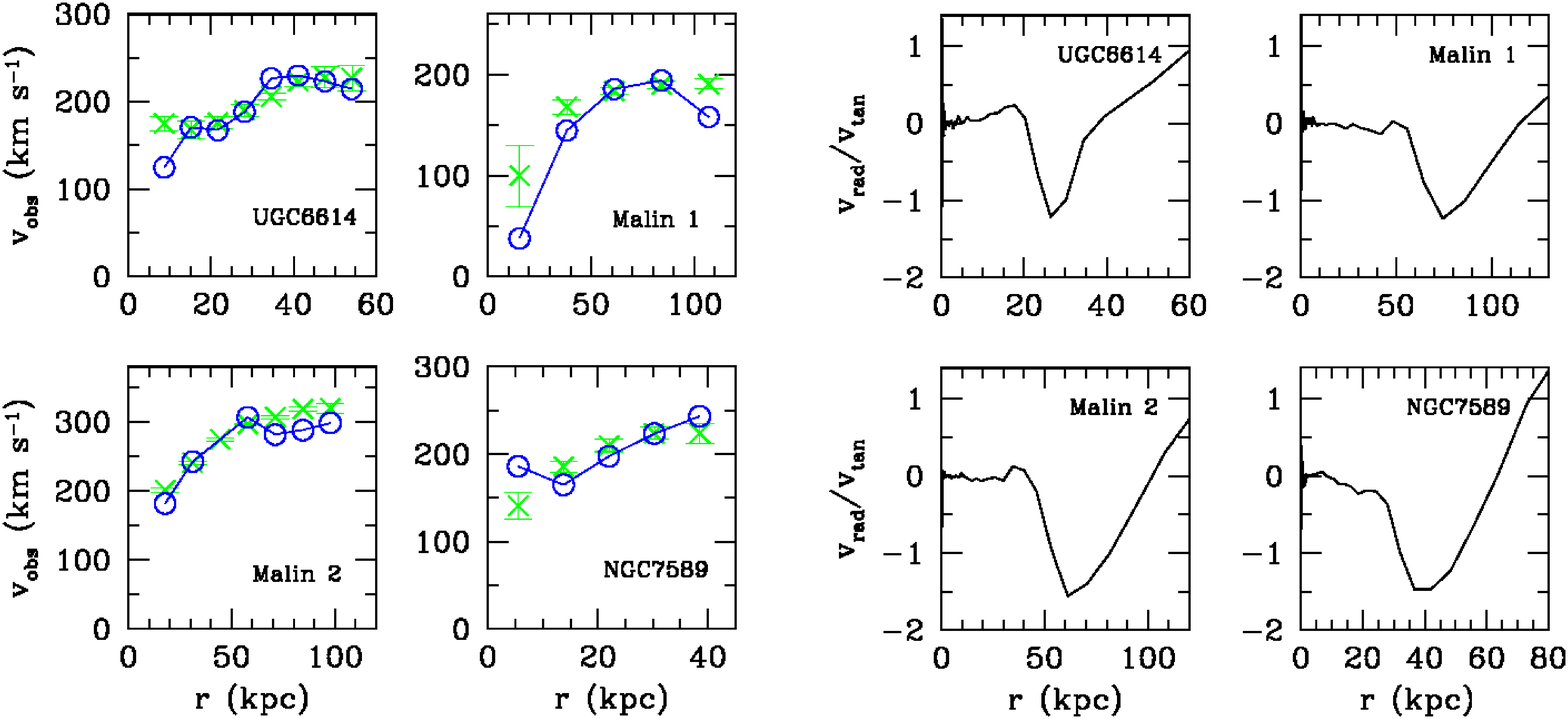} 
 \caption{
Left-hand panel: rotation curves of the GLSB sample. Green crosses  are observational data from \cite[Pickering et al. (1997)]{pickering+97}. $1\,{}\sigma{}$ errors are shown. Blue open circles connected by the solid line are the simulations. From top to bottom and from left to right: UGC6614 (run C of \cite[Mapelli et al. 2008b]{Mapelli+08b}  at $t=0.5$ Gyr), Malin 1 (run C at $t=1.4$ Gyr), Malin 2 (run B at $t=1.0$ Gyr), NGC7589 (run A at $t=0.5$ Gyr). 
Right-hand panel: Ratio between radial ($v_{\rm rad}$) and tangential velocity ($v_{\rm tan}$) of gas in the simulations. The simulations are the same as in left-hand panel.
}
   \label{fig3}
\end{center}
\end{figure}

\section{Methods}
We simulate galaxy interactions which lead to the formation of a collisional ring galaxy. The details about the simulations are reported in~\cite[Mapelli et al. (2008a)]{Mapelli+08a} and in ~\cite[Mapelli et al. (2008b)]{Mapelli+08b}. Here we remind that the simulations have been done with the N-body-SPH code GASOLINE (\cite[Wadsley et al. 2004]{wadsley+04}). Both the target and the intruder galaxy have a \cite[Navarro, Frenk \& White (1996, NFW)]{NFW} dark matter halo. The target galaxy has a stellar and gaseous exponential disc and a stellar bulge, and the ratio between the mass of the target and that of the intruder is approximately 2.

\section{Results}
The target galaxy develops a well-defined ring already $\sim{}100$ Myr after the interaction (see left-hand panel of Figure~1). The morphology of the simulated ring galaxy matches many of the properties of the Cartwheel galaxy, including the `spokes' (see the appendix of \cite[Mapelli et al. 2008a]{Mapelli+08a}). The simulation also matches the surface brightness profile of stars in Cartwheel (\cite[Higdon 1995]{higdon95}).

We continue the simulation till $\sim{}1.5$ Gyr after the interaction. We note that the ring-galaxy phase is quite short-lived: already $\sim{}$0.5 Gyr after the interaction, the ring has propagated up to $\sim{}70-90$ kpc and its surface density has significantly lowered. At the same time, the stellar disc has became extraordinarily extended and flat.
At $\gtrsim{}1$ Gyr after the encounter (right-hand panel of Figure~1) the surface density of the ring is $\sim{}2$ orders of magnitude lower than in the 'Cartwheel phase', and is almost comparable with the surrounding density. The stellar disc now extends up to  $100-130$ kpc, showing a flat surface density (in a logarithmic scale). Such flat and huge discs have been observed only in GLSBs.

We thus compare the properties of our simulated galaxies with these of the observed GLSBs. In particular, we consider a sample of four GLSBs, which have been deeply studied: UGC~6614, Malin~1, Malin~2 and NGC~7589. By the means of the package TIPSY ({\tt http://www-hpcc.astro.washington.edu/tools/tipsy/tipsy.html}) we associate to the surface density of the simulated galaxies a surface brightness. We then compare the simulated surface brightness profiles (for different runs at different times) with the observed surface brightness profiles of the four galaxies. Figure~2 shows the best-matches between observations and data (see \cite[Mapelli et al. 2008b]{Mapelli+08b} for details). The simulations match quite well the observations. We also extract the star formation histories of the simulated galaxies, and we find a good agreement with the observed star formation rate, when available. For example, the star formation rate inferred from observations for Malin~1 is $\approx{}0.1\,{}M_\odot{}$, whereas the simulated star formation rate is $\sim{}0.3\,{}M_\odot{}$ (\cite[Impey \& Bothun 1989]{impey89}). The simulations also match some interesting morphological features of many observed GLSBs, such as the existence of a bar (\cite[Pickering et al. 1997]{Pickering+97}).

Furthermore, we study the kinematics of the simulated galaxies. We rotate the simulated galaxy by the observed inclination angle and then we derive the velocity of gas particles along the line-of-sight, in order to produce simulated HI spectra. The results (shown in \cite[Mapelli et al. 2008b]{Mapelli+08b}) match quite well the data by \cite[Pickering et al. (1997)]{Pickering+97}. With a similar technique we also derive the rotation curves of the simulated galaxy. In particular, we  rotate the simulated galaxy by the observed inclination angle, we divide the galaxy into concentric annuli and for each of them we calculate the average local velocity along the line-of-sight. The results are shown in \cite[Mapelli et al. (2008b)]{Mapelli+08b} and in the left-hand panel of Figure 3. We make this kind of plots, instead than simply show the circular velocity, because we want to do something as similar as possible to what the observers do. We stress that the rotation curves that we obtain can strongly deviate from the circular velocity, as the simulated galaxies have strong non-circular motions. In fact, with the method that we adopt (and that is adopted in \cite[Pickering et al. 1997]{Pickering+97}) the entire velocity along the line-of-sight is considered and there is no way of distinguishing between circular and radial motions. Thus, the rotation curves shown in  left-hand panel of Figure 3 generally overestimate the velocity with respect to the circular velocity. This solves the apparent angular momentum discrepancy between the circular velocities of ring galaxies and of GLSBs. We also note that \cite[Pickering et al. (1997)]{Pickering+97} admit the existence of strong non-circular motions in the GLSBs they analyze and especially in  Malin~1. 

Thus, it may be important to re-analyze the existing data or to take new data and to perform a new analysis, in order to measure non-circular motions. In our simulations we have information about non-circular motions, and we can give predictions that  may be confirmed (or rejected) by future observations. In the right-hand panel of Figure 3 the ratio between radial (v$_{\rm rad}$) and tangential (v$_{\rm tan}$) velocity is shown, as a function of radius. In all the four simulated galaxies there is a clear trend: in the inner region v$_{\rm rad}$ is negligible with respect with v$_{\rm tan}$, because the central part of the galaxy is regularly rotating. In the peripheral regions v$_{\rm rad}$ is similar to v$_{\rm tan}$, because some of the ejected matter is rapidly falling back to the centre and some is still expanding in the very outer part of the ring.

\section{Conclusions}
In summary, our simulations show that collisional ring galaxies in the late stages of their evolution ($\sim{}0.5-1.5$ Gyr) match many properties of the GLSBs (surface brightness profile, HI spectrum, rotation curves, etc.). This is an interesting result, as the origin of GLSBs is an open issue, so far. In particular, the proposed scenario allows to explain the origin of GLSBs within the current cosmological model. Further theoretical and especially observational tests are required, in order to confirm this model of GLSB formation. First, kinematic data of GLSBs should be re-analyzed in order to find possible radial motions. Second,  it would be interesting to search for galaxies which are at the intermediate stage between ring galaxies and GLSBs. A possible example is UGC~6614, which is considered a GLSB, but shows a ring-like feature at $\sim{}10-20$ kpc from the centre. In our model this ring-like feature may be explained with the secondary ring produced by the galaxy interaction. Another interesting object is UGC~7069, which looks like a typical ring galaxy, but has a huge radius ($\sim{}50-60$ kpc, \cite[Ghosh \&{} Mapelli 2008]{ghosh08}). These peculiar objects deserve further studies.

\end{document}